# Directional control of weakly localized Raman from a random network of fractal nanowires

*Maria J. Lo Faro* [1,2], *Giovanna Ruello* [3], *Antonio A. Leonardi* [1,2,3], *Dario Morganti* [1,3], *Alessia Irrera* [3], *Francesco Priolo* [1], *Sylvain Gigan* [4], *Giorgio Volpe* [3,5]* and *Barbara Fazio* [3]*

[1] Dipartimento di Fisica e Astronomia, Università di Catania, via S. Sofia, 64, 95123 Catania, Italy;

[2] CNR-IMM, Istituto per la Microelettronica e Microsistemi, Via Santa Sofia 64, 95123, Catania, Italy;

[3] CNR-IPCF, viale F. Stagno d'Alcontres 37, Faro Superiore, 98158 Messina, Italy;

[4] Laboratoire Kastler Brossel, ENS-Université PSL, CNRS, Sorbonne Université, Collège de France, 24 rue Lhomond, 75005 Paris, France;

[5] Department of Chemistry, University College London, 20 Gordon Street, London WC1H 0AJ, UK.

**Corresponding Authors**

*Giorgio Volpe (g.volpe@ucl.ac.uk), *Barbara Fazio (barbara.fazio@cnr.it)

**Abstract**

Disordered optical media are an emerging class of materials capable of strongly scattering light. Their study is relevant to investigate transport phenomena and for applications in imaging, sensing and energy storage. While such materials can be used to generate coherent light, their directional emission is typically hampered by their very multiple scattering nature. Here, we tune the out-of-plane directionality of coherent Raman light scattered by a fractal network of silicon nanowires. By visualizing Rayleigh scattering, photoluminescence and weakly localized Raman light from the random network of nanowires via real-space microscopy and Fourier imaging, we gain insight on the light transport mechanisms responsible for the material's inelastic coherent signal and for its directionality. The possibility of visualizing and manipulating directional coherent light in such networks of nanowires opens venues for fundamental studies of light propagation in disordered media as well as for the development of next generation optical devices based on disordered structures, inclusive of sensors, light sources and optical switches.

The generation and control of coherent light in optical materials plays a crucial role in different branches of science and technology[1-3]. It is particularly important for fundamental physics, such as quantum physics and astronomy[4-6], as well as for applied physics with applications ranging from communication[7,8] to holography[9], and medical imaging[10,11]. Directional coherent light signals can be generated by standard cavity-based lasers[12] or by well-fabricated ordered structures, such as dielectric nanoantennas arrays[13], photonic crystals[14], waveguide systems[15,16], and plasmonic antennas[17,18]. Recently, disordered materials have emerged as an inexpensive and easy-to-fabricate alternative to these devices for the generation of coherent light[19-21], often leading to novel (and, at times, superior) optical performances to those offered by ordered photonic structures[22,23].

However, the random morphology of disordered media is generally incompatible with the generation of directional coherent signals. Speckle patterns, arising from the mutual interference of randomly scattered waves, are typical manifestation of this shortcoming of random media as the multiply scattered light leaves the optical material in random directions[3,24]. Similarly, in random lasers[25], the radiation amplified by strong multiple scattering is randomly emitted in space, unlike standard lasers where light is amplified along the axis of a cavity. Moreover, due to their multimode operating principle, random lasers typically have no control on spectral characteristics and show unpredictable lasing frequencies[26]. So far, a certain degree of control over both modal properties and spatial propagation of coherent signals in disordered materials has been demonstrated via wavefront shaping techniques only[27-29].

Directional emission of coherent light in disordered optical media is typically a very weak phenomenon, and it can only be observed as a result of averaging over many different realizations of disorder. This is, for example, the case for coherent backscattering (CBS) in random media[30-32]. In CBS, coherent light waves interfere in reciprocal paths, thus giving rise to an angular emission featuring a cone of coherently enhanced light at the exact backscattering direction. The extraction of the CBS cone from a random optical medium therefore relies on averaging different speckle realizations either by measuring the angular dependence[33] or by performing a Fourier transform[34] of the light intensity backscattered by the sample.

Here, we report the visualization and directional emission of unimodal Raman light from a fractal network of quantum-confined Si nanowires (Si NWs) without having to resort to complex external light modulation techniques. By performing real-space imaging of the inelastic light scattered by the nanostructures, we first visualize the strong scattering capability of the fractal network where photon diffusion is reduced, and light undergoes a weak localization phenomenon as measured by its localization length. By performing Fourier imaging, we then show the possibility of tuning the directionality of the weakly localized coherent Raman light from the same network of nanowires.

Practically, we fabricated disordered fractal arrays of vertically aligned Si nanowires on silicon wafer by metal-assisted chemical etching (see Supporting Section S1)[35]. The end result is a very dense vertical array of $10^{11}$-$10^{12}$ NWs cm$^{-2}$ arranged according to a random fractal planar pattern (Fig. S1). This fabrication method allows us to obtain NWs with tunable lengths from 100 nm up to a few tens of μm, and with ultrathin diameters of 7 nm on average[36]. Such a small diameter allows for the quantum confinement of charge carriers across the radial direction of the wires, thus leading to a bright room-temperature photoluminescence in the visible range[37]. Moreover, these nanowire arrays support an efficient Raman signal, which presents an asymmetric peak in agreement with the phonon quantum confinement model of Si nanostructures[37,38]. Indeed, while very low reflections are expected for this type of Si NW arrays, where the incident light resonantly bounces in a random walk across the plane of the 2D fractal, the inelastic signals (due to light-matter interactions in the sample) propagate in all directions and are very intense, as shown by the spectra in Fig. 1. In particular, the high efficiency of the Raman signal is due to the strong multiple scattering of the incident (Rayleigh) radiation within the sample, which is resonantly amplified by the underlying planar fractal morphology in the wide range of wavelengths matching its structural heterogeneities (see Supporting Section S1 and Fig. S1)[39,40]. The resonant conditions of the multiply scattered light lead to it being strongly trapped by the random array. This enhances absorption by the nanowires, ultimately leading to the excitation of a really bright photoluminescence visible to the naked eye (Fig. 1)[22,41].

Such a system of random fractal NWs therefore offers the unique opportunity to investigate light transport under laser illumination from the comparative study of three main signals: (i) the Rayleigh (elastic) multiple scattering of the incident radiation, (ii) the strong Raman (inelastic) scattering, and (iii) the efficient photoluminescence (PL) emission.

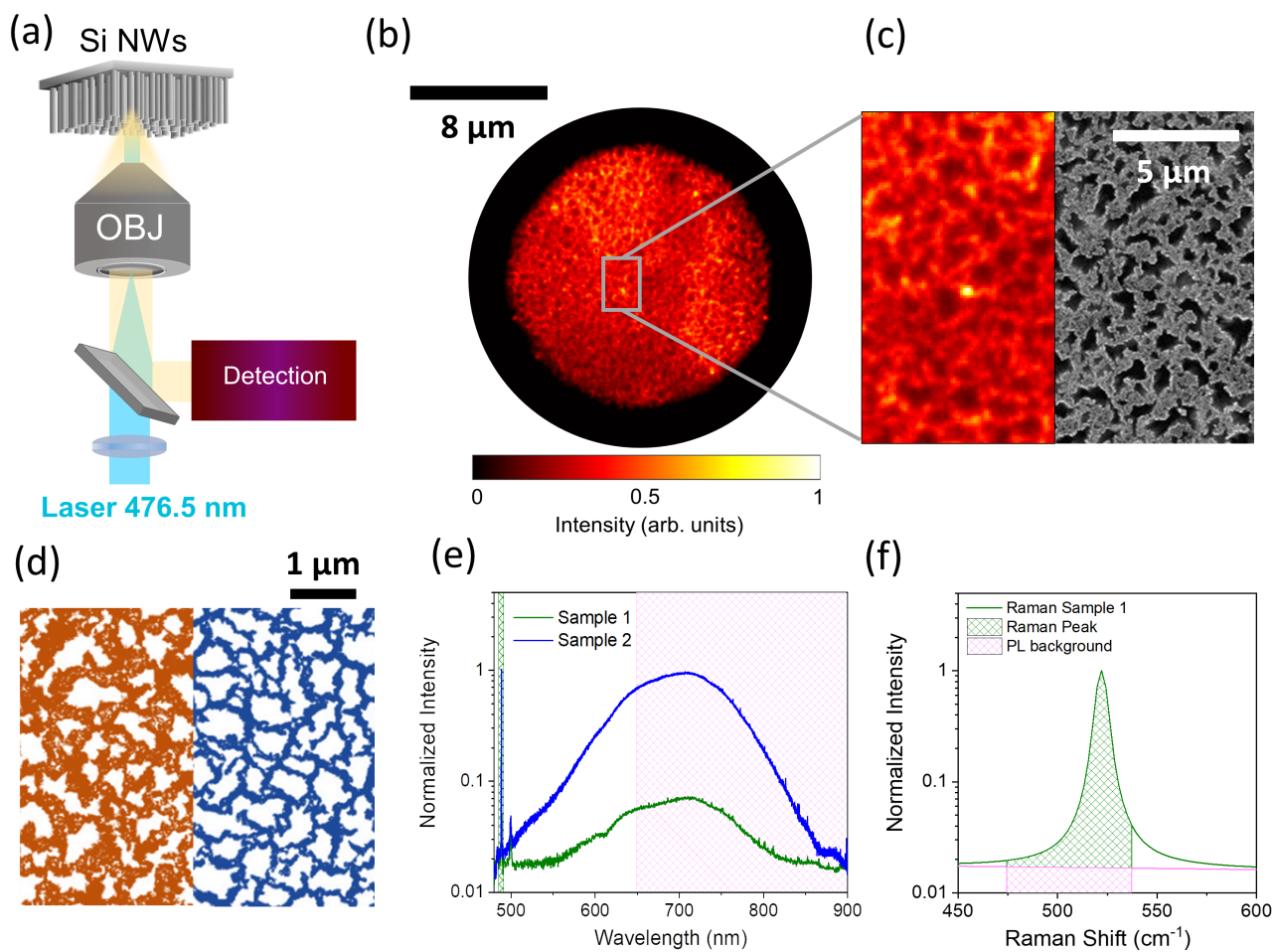

**Figure 1. Structural and optical characterization of fractal Si NW arrays.** (a) Schematics depicting the illumination of the NW samples with a laser line at 476.5 nm focused onto the back focal plane of a microscope objective (100x, N.A. = 0.9). The optical signal from the sample is collected via the same objective and redirected towards the detection system comprising a branch for real-space imaging, one for spectroscopy and one for angular detection (Fig. S2). (b) A typical bright-field optical microscopy image from a sample of Si NWs in a random fractal arrangement with 60% filling factor (Sample 1). In order to improve the contrast, the image has been acquired with a reduced field of view as obtained by reducing the aperture of the condenser diaphragm in the Köhler configuration. (c) Comparison between an enlarged detail from (b) and a scanning electron micrograph (at low magnification) of the same NW sample. The two images are reproduced with the same scalebar for direct comparison. (d) Scanning electron micrographs of Sample 1 (left) and Sample 2 (right) highlighting the different surface coverages. (e) Spectra reporting the sharp Raman signal and the broad photoluminescence band from both Si NW samples (Sample 1 and Sample 2 with stronger photoluminescence). The filled areas correspond to the spectral regions used for integrating the Raman (green) and photoluminescence signals (magenta), respectively. In particular, the photoluminescence was integrated in the 650-900nm range due to the combination of filters used to eliminate the lines corresponding to the residual laser and the Raman contribution (Supporting Section S2). (f) Detail of the Si NW 1st order Raman peak, showing the contribution of the integrated Raman signal (green) and photoluminescence background (magenta) based on the employed filters (Supporting Section S2), respectively. Note that, when excited at 476.5 nm, the PL background is very low with a negligible integrated intensity (<2%) compared to the integrated intensity of the Raman signal for both samples.

In order to investigate the optical behavior of the fractal Si NW arrays, we used a home-build microscope with three dedicated detection branches for real-space imaging, for spectroscopy and for angular detection (Fig. 1a, Fig. S2 and Supporting Section 2). In a typical experiment, we focused the external illumination provided by either a white-light lamp or a laser ($\lambda$ = 476.5 nm) on the back focal plane of a microscope objective (100x, N.A. = 0.9) to produce a nearly collimated beam (full width half maximum, FWHM ≈ 8 μm) on the air-exposed surface of the nanowires (Fig. 1a). This ensured that the illumination spot had the same phase when impinging in each point of the sample surface, which is a necessary condition for the correct evaluation of the interference effects coming from the coherent superposition of multiply scattered waves. The sample plane was then imaged on an EM-CCD camera to obtain the real-space morphology of the sample under bright-field illumination in a standard Köhler configuration to improve image contrast (see Supporting Section S2). Fig. 1b shows a typical bright-field image of a Si NW fractal array; as can be seen in the comparison between the optical and the scanning electron images reported in Fig. 1c, this bright-field image, although slightly less resolved than the corresponding SEM micrograph due to diffraction, already allows to appreciate the structural features of the fractal sample that are responsible for the optical behaviour of the material (see Supporting Section S1).

To explore the light propagation properties of such fractal random arrays, we choose to compare two samples having the same average nanowire diameter, but different surface coverages (i.e., 60% and 45% of the area under study, as shown in Fig. 1d, on all length scales). We will refer to the two samples as Sample 1 and Sample 2, respectively. The fractal dimensions ($D_F$) of the two samples are close to each other ($D_F$ = 1.87 for Sample 1 and $D_F$ = 1.81 for Sample 2), being the small variation dependent on the difference in filling factor. More importantly, the distributions of their lacunarity ($\Lambda$) are also similar, i.e. the distributions quantifying the dependance of the alternation of full and empty spaces with the length scale in the fractal structure (see Supporting Section S1 and Fig. S1). The material inhomogeneity is strongly related to the fluctuations of the refractive index in the medium, and thus determines its scattering properties. Scattering is indeed the strongest at the propagating effective wavelength $\lambda/n_{eff}$ that matches the characteristic length where $\Lambda$ assumes its peak value (here $n_{eff}$ is the refractive index of the effective medium, calculated as explained in Supporting Section S1). Due to this similarity in fractal dimension and lacunarity, we can expect that the light scattering behaviors of the two samples to be similar too. The difference in filling factor (FF) between the two samples will nonetheless influence their absorption properties, which are responsible for both the PL signal and the extinction of the light propagating through the medium. To directly compare the optical absorption of the two samples, we can introduce the inelastic scattering length $\ell_i = (\alpha \cdot Si\%)^{-1}$ (where $\alpha$ is the Si absorption coefficient at 476.5 nm), defined as the distance travelled by light before its intensity is reduced by a factor $1/e$ due to absorption[42]. The effective media are made of crystalline silicon in the nanowires' core, native silicon dioxide in the nanowires' outer shell and air voids in the gaps separating the nanowires. They are organized in finite random fractal arrays over three length scales, that extend from few tens of nanometers up to few micrometers, with the biggest air voids being around 1 μm[22,41]. Based on this, we estimated that Sample 1 has a shorter $\ell_i$ (about 2.2 times shorter) than Sample 2 due to the higher filling factor (Table 1). As a consequence, we can expect extinction and absorption effects in Sample 2 to take place over approximately double the distance than in Sample 1. Moreover, Sample 2's longer nanowires (4 μm instead of 2 μm) present a larger number of emitters for each excited wire, thus leading to a higher photoluminescence intensity.

Beyond the acquisition of real-space images, our optical setup also allows us to perform the spectral analysis of the nanowire emission through a multimode optical fiber coupled to a spectrometer (Supporting Section S2 and Fig. S2). The entire emission spectrum of our Si nanowires under illumination with laser light at 476.5 nm is reported in Fig. 1e for both samples. The photoluminescence emission (broad peak in the 500-850nm range and centered at 700 nm) is really bright, especially for Sample 2, due to its longer wires. Both spectra also present impressively high and sharp scattered Raman peaks at 488.6 nm due to the 1st order Raman scattering of the Si-Si stretching mode (corresponding to 520 cm$^{-1}$ of Raman shift) (Fig. 1f). This signal is indeed enhanced by the strong multiply scattering properties of the arrays due to their fractal nature[22].

| Si NWs | Fill factor (FF) | NWs Length (µm) | Refractive index of the effective medium $n_{eff}$ | Inelastic scattering length $\ell_i$ (µm) | PL transverse length $\sigma_{PL}$ (µm) | Raman transverse localization length $\xi_{loc}$ (µm) | Ratio $\sigma_{PL}/\xi_{loc}$ |
|---|---|---|---|---|---|---|---|
| Sample 1 | 60 ± 2% | 2 ± 0.5 | (at $\lambda_{exc}$= 476.5 nm)  1.39 ± 0.14 | 4.2 ± 0.1 | 11.5 ± 2.9 | 4.6 ± 0.7 | 2.5 ± 1.4 |
|  |  |  | (at $\lambda_{Ram}$= 488.6 nm) 1.38 ± 0.14 | 4.9 ± 0.1 |  |  |  |
|  |  |  | (at $\lambda_{PL}$= 700 nm)  1.32 ± 0.13 | 45.8 ± 0.1 |  |  |  |
| Sample 2 | 45 ± 2% | 4 ± 0.5 | (at $\lambda_{exc}$= 476.5 nm)  1.17 ± 0.12 | 9.3 ± 0.2 | 16.8 ± 4.4 | 5.5 ± 1.1 | 3.1 ± 1.5 |
|  |  |  | (at $\lambda_{Ram}$= 488.6 nm) 1.17 ± 0.12 | 10.9 ± 0.2 |  |  |  |
|  |  |  | (at $\lambda_{PL}$= 700 nm)  1.14 ± 0.11 | 101.7 ± 2.0 |  |  |  |

**Table 1. Structural and optical parameters for the fractal Si NWs arrays.** The effective refractive index values are obtained by the Bruggeman mixing rule (see Supporting Section S1.4). $\sigma_{PL}$ and $\xi_{loc}$ are average values obtained from the analysis performed on five different images for each sample.

Furthermore, with the correct combination of optical filters (see Supporting Section S2), we can image the three signals independently (Rayleigh scattering, photoluminescence emission or Raman scattering) from the same area in both samples. Typical intensity distributions of the Rayleigh, photoluminescence and Raman signals at the air-exposed interface of the Si NWs fractals are shown in Fig. 2 for Sample 1 and Sample 2. All images were acquired by using circularly polarized light with a helicity conserving polarization channel (HCC) configuration (Supporting Section S2 and Fig. S2), so to select multiply scattered light only. Indeed, the HCC contribution to the signal intensity has the dual advantage of preserving the coherence effects emerging from multiply scattering paths while intrinsically removing those due to single scattering events or direct reflections from the illumination beam. This choice is mandatory for the Rayleigh scattered radiation, therefore, for consistency, we adopted it as a standard polarization configuration for all signals.

The three optical signals in Fig. 2 provide complementary information about the light transport properties of the underlying structure (Fig. 2a-f for Sample 1 and Fig. 2g-l for Sample 2). First, as shown in Fig. 2a,g, the Rayleigh signals of the two samples are reflected and scattered from the Si nanowires as well-formed speckle patterns within the area of illumination, which are indeed characterized by a typical negative exponential intensity distribution (Fig. S3). Moving away from the illumination area (white dashed line in Fig. 2a,g), we however notice a drastic decrease in intensity until the Rayleigh light is totally extinguished, as also confirmed by the intensity profile in Fig. 2d,j for Sample 1 and 2, respectively. In fact, both the texturing of the Si NWs array and the roughness of the etched Si substrate interface at the bottom of the sample (intrinsically caused by the nanowire fabrication process) quickly couple the scattered radiation in the sample plane, where the incident light is trapped due to the multiple scattering processes resonantly amplified by the sample 2D fractality. Because of this coupling, light can propagate (minimally affected by reflections or single scattering processes) across the planar structure and out of the excitation area, until fully absorbed by the material. Note that, in order to better evaluate and compare the propagation of signals at different wavelengths through the fractal array of nanowires, we reported all profiles in Fig. 2d-f and j-l as a function of the effective length $\ell_{eff} = \ell/n_{eff}$ of light propagation from the point of entrance within the effective medium characterized by the wavelength-dependent $n_{eff}$ (Table 1).

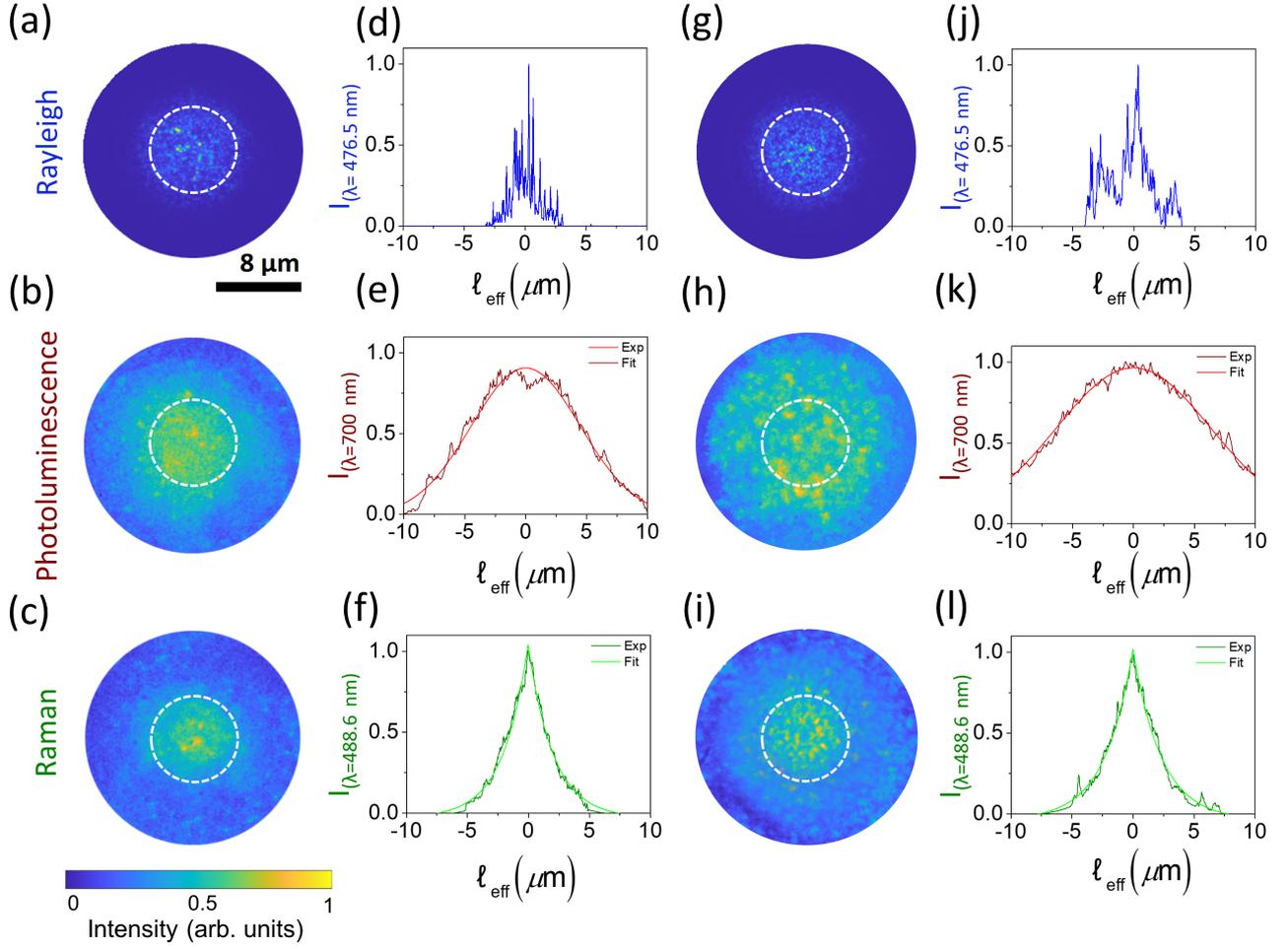

**Figure 2. Real-space optical microscopy of Rayleigh, photoluminescence and Raman light propagation in the nanowires.** (a-c and g-i) Optical images for the (a,g) Rayleigh, (b,h) photoluminescence, and (c,i) Raman signals from (a-c) Sample 1 and (g-i) Sample 2 acquired in the corresponding spectral regions of 476.5 nm, 650-900 nm, and 487-489 nm, respectively. The white dashed circles correspond to the 8-μm area of the sample directly illuminated by the collimated laser beam used for excitation (FWHM ≈ 8 μm). (d-f and j-l) Intensity profiles for (d,j) the Rayleigh signal along the diameter, (e,k) the photoluminescence and (f,l) the Raman signals as a function of the effective length $\ell_{eff} = \ell/n_{eff}$, where $\ell$ is the distance from the center of the beam scaled for the refractive index of the light in the effective medium $n_{eff}$ (Table 1). The photoluminescence and Raman profiles are averages along the angular coordinate (see Supporting Section S2) and are fitted with the characteristic (e,k) diffusive Gaussian shape and (f,l) localization exponential function, respectively.

Conversely, the images of the vast photoluminescence emission (Fig. 2b,h) visibly show a less contrasted speckle intensity distribution over a larger area with respect to the laser spot.

Photoluminescence is typically excited where there is absorption of the excitation, and, differently from the Rayleigh case (Fig. 2a,g), its image clearly provides an indirect access to the extent of the in-plane propagation of the excitation light within the material. In fact, each excited Si nanowire radiates photoluminescence in all directions in a wavelength range where the absorption by the nanowires is weaker than at the wavelength of the excitation beam, thus a significant out-of-plane photoluminescence can be collected away from the area of excitation. As confirmed by the profiles in Fig. 2e,k, being photoluminescence incoherent, its propagation in the random medium follows a diffusive regime characterized by a Gaussian intensity profile[43], $I \propto e^{-2\ell_{\text{eff}}^2/\sigma_{\text{PL}}^2}$. Here, since this quantity concerns the photoluminescence signal, the effective length $\ell_{\text{eff}}$ is scaled for the effective refractive index calculated at the wavelength where the photoluminescence is the brightest (700 nm), while $\sigma_{\text{PL}}$ is the distance along the planar structure (from the center of the exciting beam) at which the photoluminescence intensity reaches $1/e^2$ of its maximum value. Comparing the two samples, we can see how Sample 2 (Fig. 2k and Table 1) shows a photoluminescence emission from a wider area ($\approx$ 6 times the excitation area) and, consequently, a larger $\sigma_{\text{PL}}$ than Sample 1 (Fig. 2e and Table 1), whose photoluminescence emission already comes from an area $\approx$3 times bigger than the excitation area. Note that in order to measure the emission area of the photoluminescence for the two samples we took into account the FWHM of their PL intensity as obtained by the relation $\text{FWHM} = \sigma_{\text{PL}}\sqrt{2\ln 2}$ (13.54 μm and 19.78 μm for Sample 1 and 2, respectively). This is consistent with a lower absorption in Sample 2 as testified by its larger $\ell_i$, which, as noted earlier, allows for an extension of in-plane propagation of the excitation light over a much larger area than in Sample 1 (Table 1).

Finally, Fig. 2c and i show images of the Raman signal scattered from both samples, which appears more contrasted than the photoluminescence (Fig. 2b,h) and more localized than the illumination area. The presence of a weak localization in the Raman scattered light is confirmed by the negative exponential-like intensity profiles $I \propto e^{-2|\ell_{\text{eff}}|/\xi_{\text{loc}}}$ in Fig. 2f and l, where $\xi_{\text{loc}}$ is the transverse localization length (i.e. the distance where the diffusion of coherent light in the sample plane is arrested by its disorder)[43,44]. Here, $\ell_{eff}$ is obtained by scaling the propagation distance by the effective refractive index of the medium at the wavelength of Raman light (488.6 nm). Both samples have similar $\xi_{loc}$ values within the experimental error (Table 1), thus highlighting a similar scattering behaviour and confirming the scenario deduced by the analyses of the samples' lacunarities (Fig. S1). Importantly, the analysis of this real-space images gives us direct information about the localization length in our random material without the need to average over several realizations of disordered, as typically done in other random media[43]. The direct determination of the weak localization length in real space is possible since no Raman speckles are observed, as the acquisition time (tens of seconds in our experiments) is much longer than the coherence time (few picoseconds) of the phonons involved in the Raman process at each scattering site. Thus, the Raman images already are intrinsic averages over a large number of realizations of disorder, being the result of a large number of phonons produced at different times at each scattering site. The Raman signal then provides an advantage when compared to Rayleigh scattering, for which the coherence time is longer, and a localization length can only be extracted when averaging speckle patterns over hundreds different realization of disorder (i.e. over hundreds of images)[43].

The light scattering strengths of the two fractal arrays are related to their transport mean free path $\ell_t$, which is defined as the average distance travelled by light in the sample before the direction of its propagation is randomized. This important characteristic length can be generally assessed by studying the shape of the coherent backscattering cones for each sample[45,46]. In these random materials, indeed, the propagating coherent light interferes in reciprocal light paths at the backscattering direction, giving rise to a coherent backscattering cone. This is also the case for the multiply scattered Raman light[47]. Practically, this cone is the Fourier transform of the intensity distribution generated at the output sample surface by all the coherent light paths in the material. This information can therefore be obtained directly by imaging the momentum space of the sample to extract Fourier-space images. As remarked earlier in the text, unlike the elastic scattering case, the direct imaging of the cone is only possible thanks to the short coherence time of Raman light. As reported in Fig. 3, while the real-space HCC image of the Rayleigh scattered signal (Fig. 3a) still presents the expected speckle pattern in Fourier (Fig. 3b), the real-space Raman image (Fig. 3c), acquired at the same point in the sample and with the same polarization conditions, does not show a speckle pattern and presents the typical enhanced coherent signature of the nanowire Raman intensity at the exact backscattering direction (0°) in Fourier (Fig. 3d).

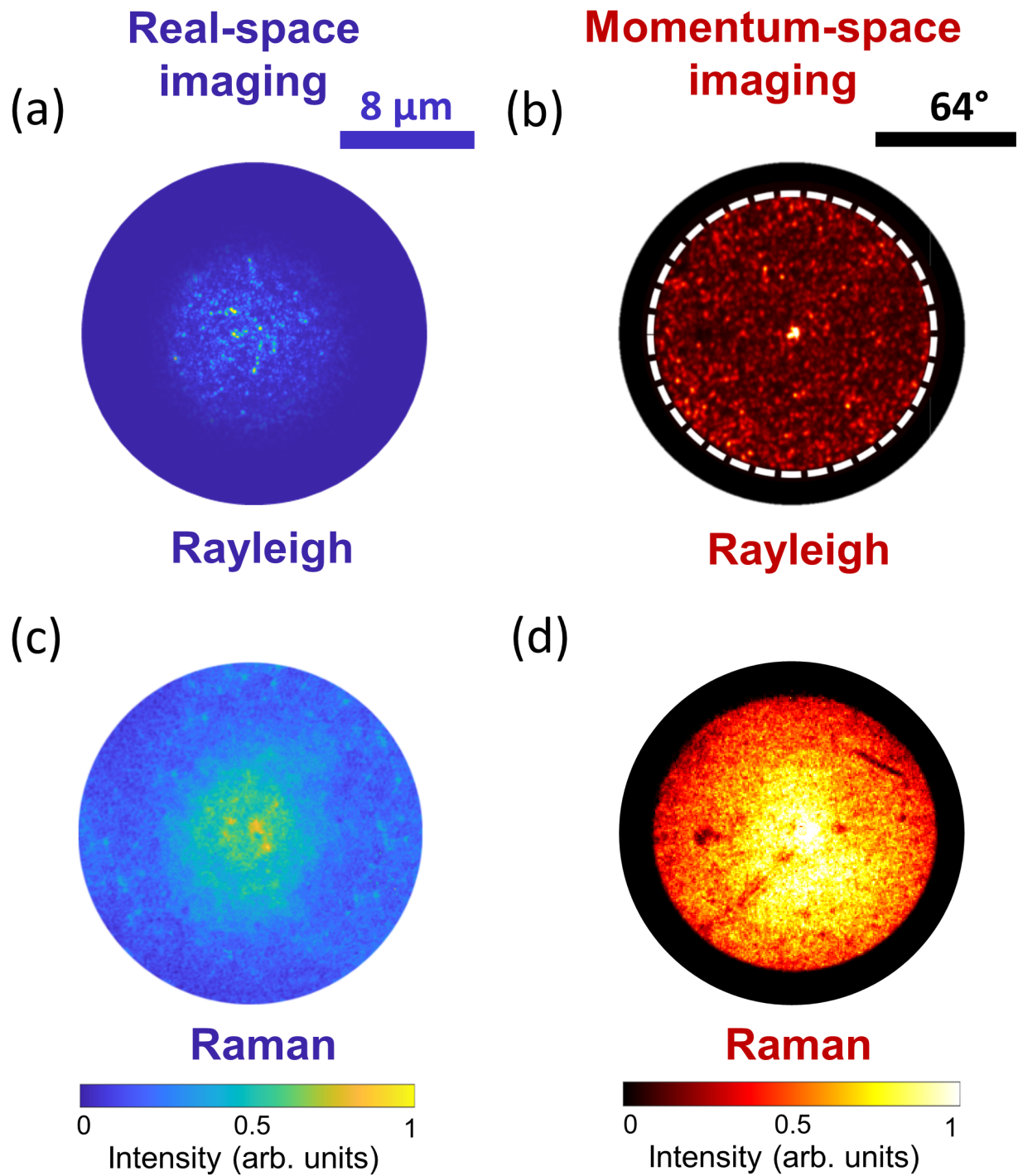

**Figure 3. Fourier microscopy and coherent backscattering visualization.** Images of (a-b) Rayleigh scattering and (c-d) Raman scattering in (a,c) real space and (b,d) Fourier (momentum space). The white

dashed circle in (b) highlights the angle corresponding to the numerical aperture (64°) limiting collection through the 100x objective (NA = 0.9) in our setup. All images have been acquired by exciting the Si NWs with the 476.5 nm laser line in the helicity conserving polarization (HCC) channel (see Supporting Sections S2).

Fig. 4 shows the experimental Raman coherent backscattering cones (dots) extracted from the momentum-space images for both Sample 1 and Sample 2 (see Fig. S4 and Supporting Section S4) and reported as a function of the detection angle θ. The cones were fitted to the theoretical expression for the Raman coherent backscattering (RCBS) given by $I(\psi) = 1 + \frac{(E_{exp} - 1)}{(E_{Raman} - 1)} \frac{\gamma_C}{\gamma_L}$, where ψ is the angle between the directions of the scattered and the incident radiations (ψ is always 0° at the exact backscattering direction independently of the direction of excitation), $E_{exp}$ and $E_{Raman}$ represent the experimental (measured) and theoretical enhancement factors of the Raman cone[47], while $\gamma_C(\ell_i, \ell_t, \ell_{dl})$ and $\gamma_L(\ell_i, \ell_t)$ are bistatic coefficients or, in other words, the expressions of both coherent and incoherent intensities in terms of the scattered fluxes per solid angle and per unit of probed area[48,49]. Both bistatic coefficients are function of the inelastic scattering length $\ell_i$ and the transport mean free path $\ell_t$, while the coherent contribution to the cone intensity ($\gamma_c$) is also function of the dephasing length $\ell_{dl}$[47]. This characteristic length ($\ell_{dl}$) arises from the dephasing mechanism developed during the mixed random walks of the excitation and of the Raman radiation, and is responsible for the lowering of the Raman cone enhancement with respect to the elastic case for which the theoretical enhancement factor is 2. In the RCBS, this length (characteristic of the excited Raman mode) is given by the relation $\ell_{dl} = 2/\Delta k$, where $\Delta k = 2\pi n_{eff} |\lambda_{exc} - \lambda_{Raman}|/(\lambda_{exc} \cdot \lambda_{Raman})$ is the magnitude of the exchanged wavevector during the Raman process, and $\lambda_{exc}$ and $\lambda_{Raman}$ are the wavelength of the excitation and of the Raman scattering, respectively. The experimental enhancement factor $E_{exp}$ can be measured from the experimental cone at the exact backscattering angle ($\psi = 0°$), while the theoretical enhancement $E_{Raman}$ is defined as $E_{Raman} = 1 + \gamma_C/\gamma_L$ [47]. In general, the theoretical enhancement differs from the experimental one, both in the elastic and in the inelastic case. Indeed, $E_{exp}$ (Raman or Rayleigh) is affected by both residual photons from single scattering events and by the stray light that introduce possible errors during data acquisition. As a consequence, $E_{exp}$ appears reduced with respect to its theoretical value. Moreover, in the Raman case, the ratio $\gamma_C/\gamma_L$ is lower than 1 due to the dephasing mechanism typical of the mixed Rayleigh-Raman random walk[47]; as a consequence, $E_{Raman}$ is always lower than 2, which is the value expected for the theoretical enhancement factor of the elastic case.

When fitting the RCBS cones in Fig. 4, the only fitting parameter is $\ell_t$, as $\ell_i$ is known (Table 1) and $\ell_{d1}$ is a fixed value for each sample given by the expression above. The calculated values of $\ell_{d1}$ are reported in Table 2: although both related to the same Raman mode the small difference between the two samples ($\ell_{d1}$ is slightly larger for Sample 2) is exclusively due to the difference in filling factor through $n_{eff}$. The obtained cone shapes appear large, which points to short transport mean free paths $\ell_t$ and, consequently, to high scattering strengths $1/k\ell_t$ (with $k = 2\pi n_{eff}/\lambda$). Moreover, within the fitting error, the fitted values of $\ell_t$ are very close to each other, confirming what we previously learnt from the real space images, i.e. that the two samples exhibit a similar behaviour in terms of scattering strength due to the similarity of their fractal dimension and lacunarity (see Supporting Section S1).

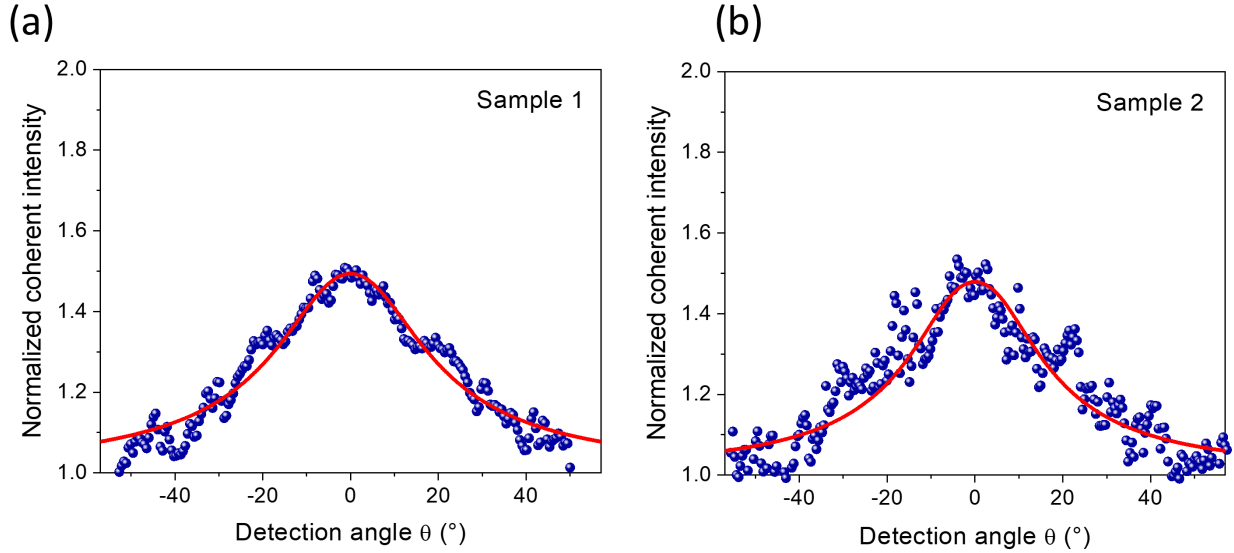

**Figure 4. Raman coherent backscattering cone.** Experimental Raman coherent backscattering cones (dots) and fitting curves (solid red lines) for (a) Sample 1 and (b) Sample 2 obtained after normalizing the HCC intensities to the respective incoherent background acquired in the cross-polarization (Vertical-Horizontal (VH)) channel (see Supporting Section S4). The red lines are the best fitting curves for the Raman cones.

| Si NWs | Dephasing length $\ell_{d1}$ (µm) | Transport mean free path $\ell_t$ (µm) | Analitical Raman transverse localization length $\xi_{loc}$ (µm) | Theoretical Enhancement Factor $E_{Raman}$ |
|---|---|---|---|---|
| Sample 1 | 4.4 ± 0.1 | 0.12 ± 0.01 | 4.1 ± 0.2 | 1.80 ± 0.02 |
| Sample 2 | 5.2 ± 0.1 | 0.14 ± 0.02 | 4.2 ± 0.2 | 1.82 ± 0.06 |

**Table 2. Raman coherent backscattering parameters for the fractal Si NW arrays.**

The role played by absorption in this process deserves an in-depth analysis, as its contribution in RCBS is really subtle. When the inelastic scattering length is large (weak absorption), the typical average dephasing between the reciprocal paths can lead to very low Raman cone intensities; conversely, when light absorption in the medium is strong, the average dephasing effect reduces and the cone is not visibility weakened[47]. Absorption therefore would become significant enough to modify the cone shape and its enhancement factor only for drastic changes in the inelastic scattering length (i.e. by at least one order of magnitude). However, since the inelastic scattering length in Sample 2 is only 2.2 times larger than in Sample 1, we do not visibly appreciate the impact of absorption on the RCBS cone, and indeed the theoretical enhancement factors $E_{Raman}$ for the two samples are found to be relatively close to each other within error bars (Table 2). Although the difference in inelastic scattering length $\ell_i$ between the two samples is not large enough to imply a consistent difference in terms of the shape of their Raman cones, it is large enough to justify the difference in the photoluminescence emission between the two sample (Fig. 1), since it strongly influences the extent of light propagation inside the media (Fig. 2).

Starting from the values of transport mean free paths obtained by fitting the RCBS cones, we are also able to calculate the transverse localization lengths $\xi_{loc} = \ell_t e^{\pi k \ell_t / 2}$ (for the 2D case) for both samples analytically[24,43]. The values for $\xi_{loc}$ are in good agreement with those previously derived by fitting the intensity distribution profiles in the real-space Raman images (Fig. 2 and Table 1).

Finally, the direct visualization of the directional Raman CBS cone from the samples (i.e. without having to average over a large number of disorder configurations) opens up new venues for utilization of random media as devices for the emission of directional coherent light. As shown in Fig. 5, when the direction of the incident radiation is changed, the directionality of the coherent Raman emission follows, being directed along the exact backscattering direction on each occasion. This is a typical fingerprint of CBS phenomena. In particular, we imaged different Raman cones as obtained by focusing the excitation beam at different points of the objective back-focal plane, thus changing its direction of propagation $\phi_{exc}$ (its *k*-vector) from $\phi_{exc} = 0°$ to $\phi_{exc} = -10°$ to $\phi_{exc} = -20°$ with respect to normal incidence on the sample surface. As a result, the directionality of the maximum intensity in the Raman cone also shifts accordingly (Fig. 5).

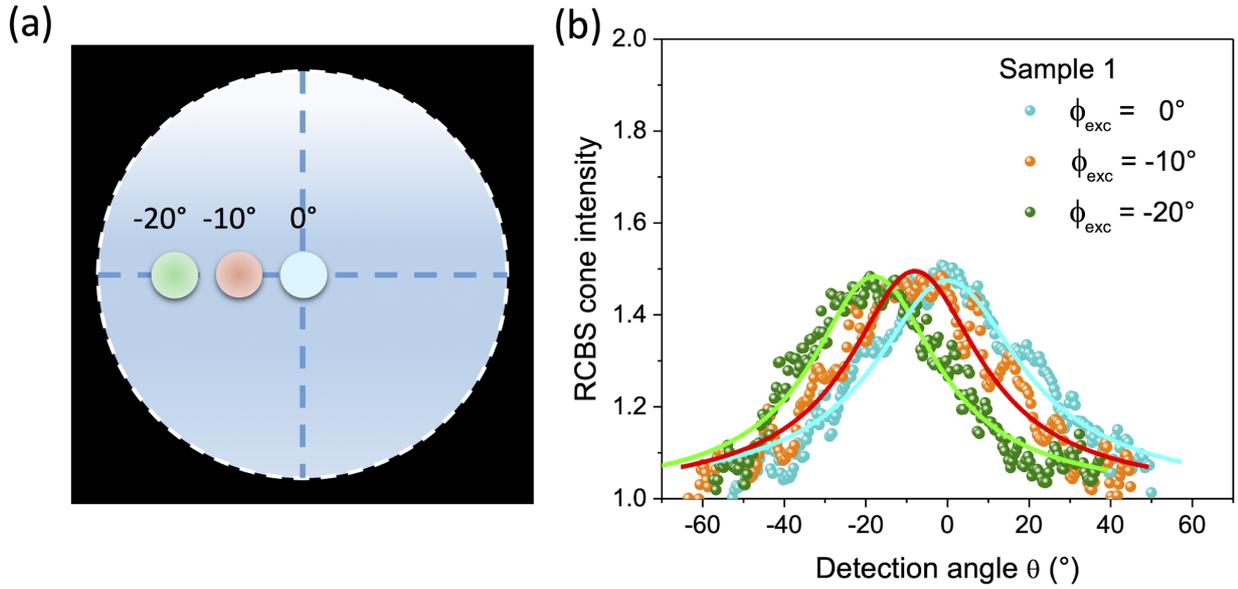

**Figure 5. Directional emission of the Raman coherent backscattering cone.** (a) Scheme depicting the focusing of the laser beam in the back-focal plane of the objective in correspondence of the following angles: $\phi_{exc} = 0°$ (blue dot), $\phi_{exc} = -10°$ (orange dot), and $\phi_{exc} = -20°$ (green dot). Here, $\phi_{exc}$ is the angle between the incident radiation and the normal to the sample surface. (c) Normalized Raman coherent backscattering cones (dots) and fitting curves (solid lines) from Sample 1 obtained at a laser incidence of $\phi_{exc} = 0°$ (blue), $\phi_{exc} = -10°$ (orange), and $\phi_{exc} = -20°$ (green). Note that, in this case, the exact backscattering angles $\psi = 0°$ coincide each time with the detection angle $\theta=0°$, $\theta=-10°$, and $\theta=-20°$.

**Conclusions**

In this work, we have simultaneously visualized the optical signals (Rayleigh, photoluminescence and Raman) coming from random fractal networks of silicon nanowires under laser excitation. These signals were visualized both in real space and in momentum space. The combined imaging permitted us to gain insight on the complex scattering and absorption properties behind the investigated materials as well as their interplay beyond what typically allowed by standard angle resolved CBS techniques only. In particular, imaging the Si NWs photoluminescence emission allowed us to directly determine the extent of the in-plane propagation of the excitation light within the disordered materials. Furthermore, a single coherent Raman image permitted us to directly probe its weak localization length, which would not be possible to extract otherwise from the elastically scattered radiation unless averaging over a large number of different disorder realizations. By imaging the process in momentum space and by taking into account the physics of Raman scattering, we were able to obtain the first direct visualization of the peculiar phenomenon of Raman coherent backscattering. So far, this occurrence has only been observed by means of angle resolved light scattering measurements. Previously, momentum-space imaging had been employed for patterning the angular distribution of typical Raman bands in 2D materials, such as graphene[50], in order to evaluate their polarization ratio or in plasmonic antennas[51-53] for estimating their degree of directional emission. In this work, we have further demonstrated the possibility of using Fourier imaging for tuning the directionality of coherent Raman light scattered by a random medium. This demonstration, coupled to the direct visualization of Raman localization, is a convenient way forward to generate coherent and directional light emission from disordered structures, which could be easily implemented in devices without the need for external light modulation techniques[54,55]. Coherent Raman radiation offers the additional advantage of fine frequency tuning based on the small energy loss (Stokes) or acquisition (anti-Stokes) typical of Raman scattering processes. Ultimately, our finding paves the way for the development of next generation inelastic coherent light sources based on random materials with out-of-plane directional coherent light control and fine wavelength tuning capabilities.

**Supporting Information:** Section S1 includes the fabrication protocol of the silicon nanowires samples (S1.1), their structural and fractal characterization (S1.2, S1.3), and the estimation of their refractive index (S1.4). Section S2 describes the experimental optical set up used for the acquisition of real-space and Fourier images. Section S3 includes the intensity distribution of the Rayleigh speckle patterns. Section S4 describes the analysis of the experimental Raman Coherent Backscattering Cone.

**Author Contributions**

G.V. and B.F. conceived the idea for the work, designed the experiments, supervised and coordinated the project. S.G., G.V. and B.F designed the experimental setup. G.R. and B.F. built the setup. M.J.L.F., D.M.

and A.I. fabricated the samples. G.R. and B.F. performed the experiments. M.J.L.F. and B.F. analysed the data. M.J.L.F., G.V. and B.F. interpreted the data. M.J.L.F., G.V. and B.F. wrote the manuscript. A.A.L, A.I., F.P. and S.G. provided comments on a draft of the manuscript. All authors read the final version of the manuscript.

**Acknowledgements**

We acknowledge D. Arigò, G. Lupò, S. Patanè, G. Spinella and S. Trusso for expert technical assistance. Giorgio Volpe, Barbara Fazio, Antonio A. Leonardi and Maria J. Lo Faro acknowledge support from the Royal Society under grant IE160225. Giovanna Ruello acknowledge support from project "STBIC" – P.O. FSE 2014/2020. Project PON ARS01_00459 ADAS is acknowledged.

**Conflict of interest:** the authors declare no conflict of interest

ABBREVIATIONS

CBS, coherent backscattering; NWs, nanowires; 2D, two-dimensional; FWHM, full width half maximum 5; EM-CCD, Electron Multiplying Charge Coupled Device; SEM, scanning electron microscopy; FF, filling factor; PL, photoluminescence; HCC, helicity conserving channel; RCBS, Raman coherent backscattering; VH, vertical-horizontal.

# SUPPORTING INFORMATION

# Directional control of weakly localized Raman from a random network of fractal nanowires

*Maria J. Lo Faro* [1,2], *Giovanna Ruello* [3], *Antonio A. Leonardi* [1,2,3], *Dario Morganti* [1,3], *Alessia Irrera* [3], *Francesco Priolo* [1], *Sylvain Gigan* [4], *Giorgio Volpe* [3,5]\* *and Barbara Fazio* [3]\*

[1] Dipartimento di Fisica e Astronomia, Università di Catania, via S. Sofia, 64, 95123 Catania, Italy; [2] CNR-IMM, Istituto per la Microelettronica e Microsistemi, Via Santa Sofia 64, 95123, Catania, Italy; [3] CNR-IPCF, viale F. Stagno d'Alcontres 37, Faro Superiore, 98158 Messina, Italy; [4] Laboratoire Kastler Brossel, ENS-Université PSL, CNRS, Sorbonne Université, Collège de France, 24 rue Lhomond, 75005 Paris, France; [5] Department of Chemistry, University College London, 20 Gordon Street, London WC1H 0AJ, UK.

**Corresponding Authors**

\*Giorgio Volpe (g.volpe@ucl.ac.uk), \*Barbara Fazio (barbara.fazio@cnr.it)

# Section S1: Fabrication protocol and characterization of silicon nanowires samples.

**S1.1: Fabrication protocol**

Fractal arrays of silicon nanowires (NWs) were obtained by metal-assisted chemical etching using a thin percolative gold layer as a precursor. Nanowires obtained by this methodology are made of a crystalline core of silicon (Si) embedded into an outer shell of silicon dioxide ($SiO_2$. After their synthesis, any remaining gold layer is chemically removed. More in detail, n-type 111-Si wafers (Siegert Wafer) were first treated with UV-ozone and then etched in an aqueous solution at 5% of HF (49% hydrofluoric acid, Sigma-Aldrich) in order to remove the native oxide layer on the surface. A 2-nm-thick discontinuous gold layer (99.99%, CinquePascal) was then deposited on the clean wafers by using an electron beam evaporator (KM500, Kenosistec) at room temperature. The sample (gold layer + Si substrate) was etched at room temperature in an $HF/H_2O_2$ aqueous solution (5:0.5 M) for the synthesis of the nanowires and then rinsed in KI solution (Sigma-Aldrich) for Au removal. In this work, two different nanowire samples (Sample 1 and Sample 2) were grown on the same type of Si substrate upon the deposition of a 2 nm-thick Au layer obtained at different deposition rates, thus realizing two slightly different fractal geometries. The different deposition rates allow for a distinct distribution of the Au atoms in slightly nonidentical percolative layers, which result in different filling factors (FF) and, hence, different effective refractive indexes ($n_{eff}$). Sample 1 (FF ≈ 60%) and Sample 2 (FF ≈ 45%) were obtained with an Au deposition rate of about 0.2 $Ås^{-1}$ and 0.8 $Ås^{-1}$, respectively. Moreover, the length of the nanowires in the two samples was also varied from 2 μm (Sample 1) to 4 μm (Sample 2) by using different etching times (12 min and 20 min, respectively) as shown in Fig. S1.

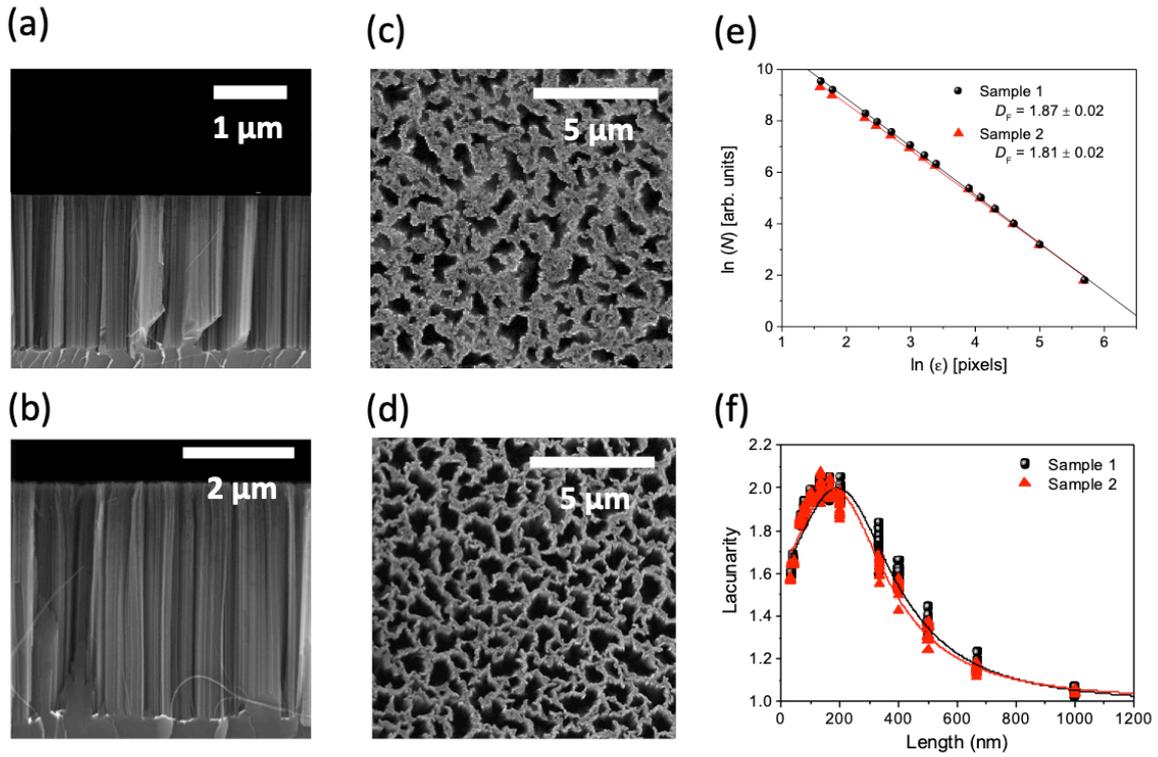

**Figure S1: Synthesis and structural characterization of silicon nanowires array.** (a-b) Cross-section and (c-d) plan-view scanning electron microscopies of the fractal array of Si nanowires for Sample 1 (a,c) and Sample 2 (b,d), respectively. (e-f) Measured (e) fractal dimension and (f) lacunarity for Sample 1 (black circles) and Sample 2 (red triangles). The solid lines (black and red) are the respective Lorentzian fitting curves.

**S1.2: Structural Characterization**

Morphological information about the nanowires was obtained using a scanning electron microscope (SUPRA, Zeiss) equipped with an InLens detector for high-resolution imaging and with an energy dispersive X-ray analyzer (EDAX) for analytical measurements. Figure S1 (a-d) shows the morphologies of two NW fractal arrays as acquired from the SEM investigations. The SEM cross-sections in Fig. S1a and b display the realization of very dense vertically aligned Si nanowires obtained by the fabrication method described in Section S1.1. The NW length increases with the etching time from 2 μm (Sample 1) up to 4 μm (Sample 2). The plan-view SEM micrographs in Fig. S1 c and d show the different surface area coverage characteristic of the two NW samples, measured by their respective filling factors (FF). The filling factor is defined as the ratio between the area covered by the nanowires and the total area of the analyzed image (FF = covered area/total area) and was measured by pixel counting using Image J software. Hence, from Fig. S1 c and d, we measured filling factors of about 60% and 45% for Sample 1 and Sample 2, respectively.

**S1.3: Fractal properties**

As previously reported, these NW arrays are fractals[1,2]. Their characteristic parameters of *fractal dimension* ($D_F$) and *lacunarity* ($\Lambda$) were measured as a function of the scale invariance of the system. The fractal dimension $D_F$ measures the complexity of the structure, and shows how the fractal geometry scales as a function of the investigated length scale[3]. The lacunarity $\Lambda$ is defined as a statistical fluctuation in the distribution (here, the normalized pixel density distribution) of the filled regions of the sample versus its empty regions measured as a function of the investigated length scale. We measured the fractal parameters of the NWs structures by using the sliding box-counting algorithm (Sbcount) implemented in the FracLac plugin of the ImageJ software[4]. Briefly, both parameters were measured by counting the number of the filled pixels ($N$) in each box of a grid of size ε superimposed onto the SEM plan-view. The process is iterated by varying the grid size (ε) from 37 nm up to 2.2 $\mu$m to achieve convergence, and repeated for five different initial grid positions in order to acquire a significant statistical ensemble. As this algorithm works on binary images, the SEM plan-view microscopies were first converted into black and white images by ImageJ software. The same magnification and image size conditions were used for all the analyzed set of images for consistency. A 50kX magnification was chosen since it allows us to simultaneously probe bigger and smaller dimension ranges, respectively well above and below hundreds of nanometers. The measured fractal dimensions and lacunarities for the two samples are reported in Fig. S1 e and f, respectively. Figure S1 e reports the average number $N$ of occupied pixels per box as a function of the box sizes ε, where a linear relation is obtained between these two values in a log-log scale. The fractal dimension $D_F$ is then obtained from the slope of the linear function fitting the pixel density statistics per box size ($D_F = \log N / \log \varepsilon$). The expected $D_F$ for fractal structures is a non-integer value and, as far as 2D fractals are concerned, it is strictly smaller than the Euclidean value of 2. Our analysis confirms that both samples of NWs are 2D fractals with random geometries with a fractal dimension of 1.87 and 1.81 for Sample 1 and Sample 2, respectively. The lacunarity was calculated instead as $\Lambda_{e.g} = 1 + \sigma^2/\mu^2$, where σ and μ are the standard deviation and mean of the pixel density distribution calculated for each set of grid dimensions ε and grid positions used to investigate the image. The lacunarity peak is associated with the maximum heterogeneity in the structure; hence it corresponds to the length scale at which the variations of the effective refractive index are maximized. When light propagates through the medium with an effective wavelength ($\lambda_{eff} = \lambda/n_{eff}$) that matches the length scale at which the lacunarity reaches its peak, the scattering inside the structure is maximized too. Figure S1 f shows that, despite a small shift in the lacunarity peak, the fluctuation of the pixel density is similar for both samples.

**S1.4: Estimation of the refractive index**

The Si NWs fractal arrays show scale invariance as attested by SEM investigations over three orders of magnification (5kX, 50kX and 500kX) and, therefore, over three length scales (Fig. S1 e and f). The three length scales extend from a few tens of nanometers up to few micrometers, being the maximal dimension of the observed air voids around 1 μm[1,3]. Considering the fractal properties of our samples, the concentration of crystalline silicon decreases as the observation scale increases. For the largest magnification (corresponding the field of view under the optical microscope in Fig. 1), the percentage of crystalline Si is as low as 10% for Sample 1 and 4.5% for Sample 2. Starting from these percentages and from the effective composition of the materials (made of crystalline silicon in the nanowires core, native silicon dioxide in the nanowires outer shell and air voids in the gaps separating the nanowires), we calculated the effective refractive index $n_{eff}$ for the two samples applying the Bruggeman mixing rule[5,6], in order to correctly interpret the light transport properties of such complex fractal media.

## Section S2. Optical set up: acquisition of real-space and Fourier images

All optical images and spectra were obtained with the homemade setup detailed in Fig. S2. Briefly, the 476.5 nm excitation from an Ar+ laser (Spectra Physics 2020) is focused on the back focal plane of a microscope objective (100x, NA = 0.9, Olympus - BX41) by a biconvex lens (L1, Thorlabs) to produce a collimated beam impinging on the sample. Bright-field Köhler illumination[7] was also implemented by using a white Xenon lamp (Osram) as an alternative source for sample imaging. For image detection, we used a scheme which includes relay optics and a Bertrand lens[8,9]. In this configuration, the objective back focal plane is projected through a beam splitter (BS, Thorlabs) and a second biconvex lens (L2, Thorlabs) to form an image of the sample at the image plane (IP). This image is then either recreated by an f-f-f'-f' configuration or Fourier-transformed in a momentum space image via an f-f-2f'-2f' configuration on an EMCCD camera (Andor). The f-f-f'-f' configuration (green dashed lines in Fig. S2) is implemented by using two biconvex lenses (L3 and L4, Thorlabs), while the f-f-2f'-2f' configuration (magenta dashed lines in Fig. S2) is implemented replacing L4 with a biconvex lens (L5, Thorlabs) having half the focal length of L4. Finally, a fiber-coupled spectrometer (LotOriel 260i) was employed for spectral analysis as depicted in Fig. S2.

After proper signal filtering, Rayleigh, Raman, and photoluminescence can therefore be observed in real space, in Fourier, or can be spectrally analyzed. A laser interference filter at 473 nm (LD01-473/10, Semrock) was consistently employed to remove the residual laser line from the detection path.

The acquisition of different optical signals was then performed employing tailored filter sets for detection. In particular, Rayleigh light was detected using an optical density filter (OD3, CVI Laser Corporation) to decrease its intensity and avoid saturating the camera. Raman and photoluminescence signals were acquired after removing the Rayleigh line with an edge laser filter (BLP01-473R-25, Semrock). For photoluminescence, two additional cooperative long-pass filters were used (LP550 + LP650, CVI Laser Corporation) for the acquisition of the NWs emission, whose complete range extends from about 500 nm to 850 nm. After the PL filter set, the spectral range is restricted to wavelengths above 650 nm. For Raman, a narrow bandpass filter at 488 ± 1 nm (LL01-488-25, Semrock) was used, as this wavelength range corresponds to the 1st order Raman mode at 520 cm$^{-1}$ of the Si-Si bond stretching (observed at 488.6 nm when excited with 476.5 nm). In addition, a short-pass filter (SP550 from Edmund Optics) was also used to remove the NWs PL emission in the red region (500 - 850 nm). We used the spectrometer to verify the effectiveness of the adopted filter sets to discriminate among the three signals and remove undesired spectral contributions (see Fig. 1 and Fig. S2).

All the real-space images were acquired under a helicity conserving channel (HCC) configuration (Fig. S2) by using in order: 1) a linear polarizer (Thorlabs) to select the input linear polarization, 2) a quarter waveplate (0$^{th}$-order λ/4 at 473 nm, Thorlabs) in order to convert the input linear polarization to circular (note that the output signal from the sample passes a second time through the same quarter waveplate), and 3) another linear polarizer (Thorlabs) in the detection part to select the linear polarization parallel to the input as to preserve helicity. This configuration allows us to exclude single scattering events, since they do not preserve helicity, and thus to acquire the coherent signal arising from multiply scattering paths only. Finally, since incoherent contributions with the same helicity can overlap to the coherent ones of the same acquired signal (e.g., in the Raman case), we also need to obtain the incoherent background and normalize the HCC intensity to it in order to obtain only the coherent contribution to the signal intensity. To obtain the incoherent background, we adopted a linear cross-polarization configuration (VH) simply by removing the quarter waveplate and rotating the output polarizer by 90° (cross-polarization).

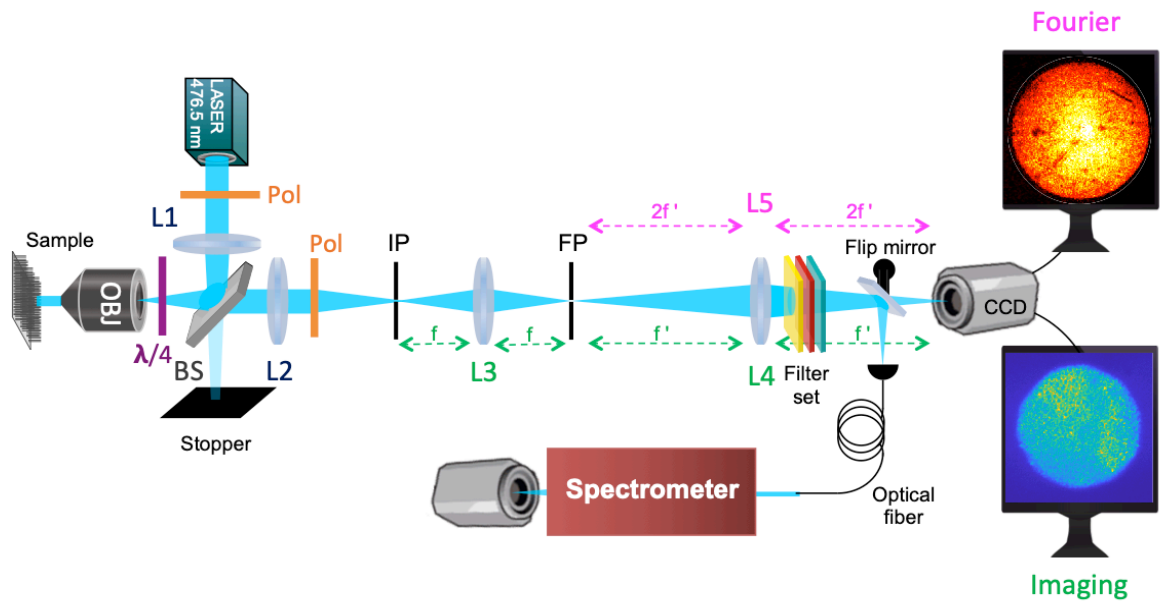

**Figure S2: Detailed schematics of the optical setup.** The 476.5 nm Ar· laser line is focused at the back focal plane of the objective (100x, NA = 0.9) with a lens L1 as to be arrive collimated to the sample. The real-space image or Fourier transform of the sample plane are then projected on an EMCCD camera through a set of lenses (L3 and L4 or L3 and L5, respectively). The use of a proper filter set allows us to acquire the optical signal in the desired spectral range as verified with a fiber-coupled spectrometer. A set of two linear polarizers (Pol), and a flip quarter-wave plate (λ/4) were used to select the proper polarization channel for the analysis of the CBS cone. The focal lengths of the lenses in the setup are: $f_{L1}$ = 30 cm; $f_{L2}$ = 30 cm; $f_{L3}$ = 25 cm; $f_{L4}$ = 20 cm; $f_{L5}$ = 10 cm.

All images from the CCD were cropped to selected active regions. First, the background was subtracted using ImageJ. Then, the images were plotted and analyzed in MATLAB to extract intensity profiles. For each image, an intensity profile was obtained as an average of 4 distinct profiles (vertical, horizontal, and two diagonal profiles). For the same sample, five different images were acquired at different points; their profiles were extracted as above and used to obtain the average profiles shown in Fig. 2.

## Section S3. Intensity distribution of Rayleigh speckle patterns.

The Rayleigh signals of both Sample 1 and Sample 2 form well-defined speckle patterns. These speckles are characterized by typical negative exponential intensity distributions[10]:

$$p(I) = \frac{1}{\langle I \rangle} e^{-I/\langle I \rangle}$$, as shown in Figure S3.

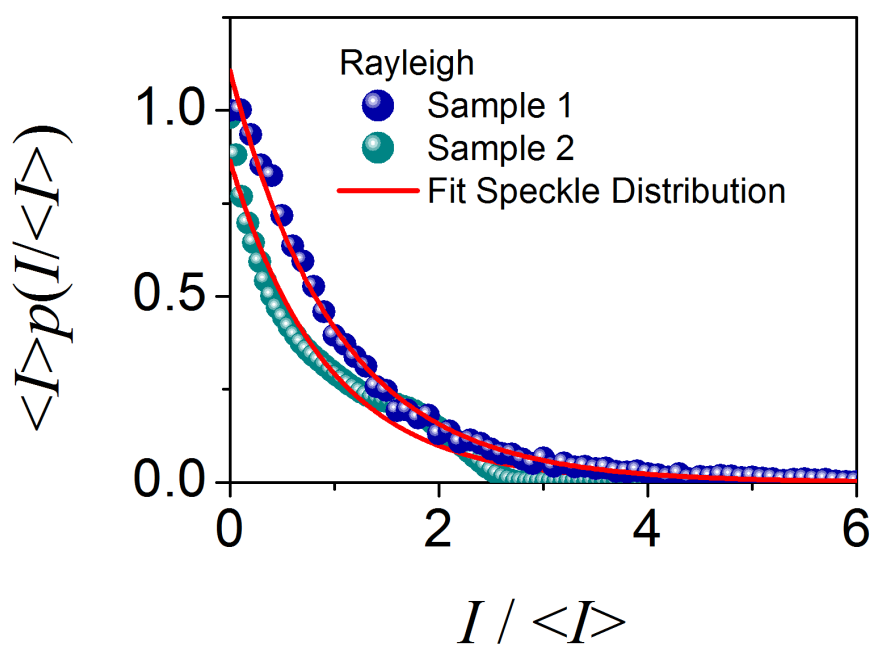

**Figure S3: Speckle pattern intensity distributions.** Intensities distributions of Rayleigh speckle patterns for Sample 1 (blue dots) and for Sample 2 (green dots). The red lines represent the negative exponential fitting curves.

## Section S4. Analysis of the Raman Coherent Backscattering Cone

In Fig. S4 we show the momentum-space images of the Raman signal for the helicity conserving polarization channel (HCC, Fig. S4a,c) and for the linear cross-polarization channel (VH, Fig. S4b,d) for both Sample 1 (Fig. S4a-b) and Sample 2 (Fig. S4c-d). In order to extract the cone shape, we first obtained an average intensity profile for each image in the HCC and VH channels. In particular, by using the plot profile function from ImageJ software (https://imagej.nih.gov/ij/), we averaged the profiles obtained along the diameters in the *x*-axis, the *y*-axis, and the two main diagonals of the circle defined by the numerical aperture of the objective. We then scaled the average intensity profiles for the HCC and VH channels in order to overlap their values at large angles (around 55°). Finally, we normalized the two profiles to one another ($I_{HCC}/I_{VH}$) in order to achieve the enhanced Raman coherent backscattering cone (RCBS) for each sample as shown in Fig. 4.

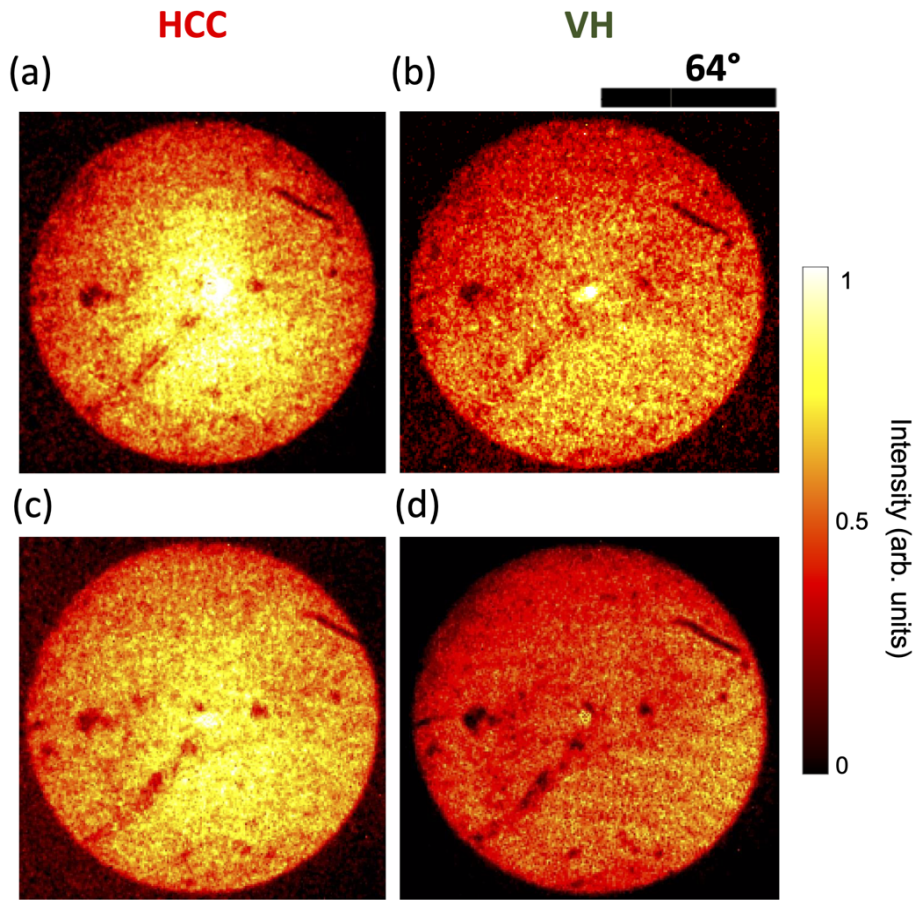

**Figure S4: Extraction of the of the Raman Coherent Backscattering Cone.** Momentum-space images of the nanowire Raman signal acquired in (a,c) the helicity conserving channel (HCC) portraying the superposition of both coherent signal and incoherent background, and (b,d) the linearly cross-polarized channel (VH) exclusively portraying the incoherent background for Sample 1 (a,b) and 2 (c,d), respectively. For each sample, the Raman coherent backscattering cone is obtained by processing the average intensity profiles extracted from these images (Fig. 4 and Supplementary Section S4).